\documentclass{aa}

\newcommand{\fe}{Fe~K$\alpha$}
\newcommand{\etal}{et al.}
\newcommand{\kev}{keV}
\usepackage{graphicx}

\begin{document}

\title{X-ray spectral diagnostics of the immediate environment of GRB~991216}
\titlerunning{X-ray spectral diagnostics of GRB~991216}

\author{David R. Ballantyne\inst{1}, Enrico Ramirez-Ruiz\inst{1},
Davide Lazzati\inst{1} \and Luigi Piro\inst{2}}
\authorrunning{Ballantyne et al.}

\offprints{D. Ballantyne}

\institute{Institute of Astronomy, University of Cambridge, Madingley
Road, CB3 0HA Cambridge, England \\
\email{drb,enrico,lazzati@ast.cam.ac.uk}
\and Istituto di Astrofisica Spaziale, CNR, Via del Fosso del 
Cavaliere 100, 00133 Roma, Italy \\
\email{piro@ias.rm.cnr.it}}

\date{}

\abstract{The recent report of iron line features in the afterglow of
the gamma-ray burst (GRB) 991216 has important implications for the
properties of the radiating material and hence the nature of the
immediate burst environment. We argue that the putative strong Fe
emission line can be attributed to the reflected emission from
Thomson-thick matter which is illuminated by a power-law
continuum. The ionization parameter of the material (i.e., the flux to
density ratio) is around $10^{3}$, resulting in a Fe K$\alpha$ line from
He-like iron. A supersolar abundance of iron is not required by the
data. Interestingly, the ionizing continuum must be harder than the
observed one.  We interpret this as due to the fact that the
observed continuum is dominated by the standard blast wave emission,
while the line is produced by reprocessing material located at much
smaller radii.
}

\maketitle

\section{Introduction}
\label{sec:int}

The detection of iron emission features in the X-ray spectra of
several $\gamma$-ray burst (GRB) afterglows (GRB~970508: Piro et
al. 1999; GRB~970228: Yoshida et al. 2001; GRB~991216: Piro et
al. 2000; GRB~000214: Antonelli et al. 2000) is of great importance to
the understanding of the nature of the burst emission and particularly
the burst progenitor (e.g., M{\'e}sz{\'a}ros \& Rees 1998; Lazzati et
al. 1999; B\"ottcher \& Fryer 2001).  All the lines have been detected
roughly one day after the burst explosion due to instrumental
limitations (the line may have been there at earlier times, but the
afterglow was not yet observed), with an equivalent width (EW) of the
order of 1~keV and a luminosity
$L_{\rm{Fe}}\approx10^{44}$~erg~s$^{-1}$.

The large EW inferred from the X-ray features favour models in which
the line is produced by reflection from a highly ionized surface,
rather than transmission. Several different emission mechanisms and
geometries have been proposed to account for the line emission, all of
which fall mainly into one of two categories: {\it geometry} dominated
(GD) models or {\it engine} dominated (ED) models. Both classes
associate the scattering medium with debris from a stellar
collapse. In the GD models, the X-rays from the burst and early
afterglow emission illuminate a mass $M_{\rm{Fe}}>0.06M_{\odot}$ of
iron material at a distance of about $10^{16}$~cm from the burst
location (Lazzati et al. 1999; Vietri et al. 2001; see also Lazzati et
al. 2001). This material may be associated with the remnant of a
supernova that occurred days to months before the burst (Vietri \&
Stella 1998). Alternatively, in the ED models, the line emission can
be attributed to the interaction of a long lasting outflow from the
central engine with the progenitor stellar envelope at distances $R
\le 10^{13}$~cm (Rees \& M{\'e}sz{\'a}ros 2000; M{\'e}sz{\'a}ros \&
Rees 2001; B\"ottcher \& Fryer 2001). In this case, only a small mass
of Fe is required and there is no need for a pre-ejected supernova
shell (i.e., the line can be explained if the burst and the star
explode simultaneously; see Woosley 1993).

There are several ways in which these two models can be distinguished.
First, the line produced by a compact reprocessor will show a higher
degree of variability. It is likely that in any ED scenario the line
intensity will decrease with time, mirroring the decay of energy
input. Due to geometrical time dilution of the line photons, it is
hard for any GD model to produce line variability on a time scale
shorter than $t_{\rm var}<R/c\approx1$~d. Unfortunately, the detected
lines were too weak to allow for a proper variability analysis, and no
firm conclusion could be made (while a hint of non-variability was
derived in GRB~000214 [Antonelli et al. 2000], the line in GRB~970508
seemed to disappear before the end of the observation [Piro et
al. 1999]). Secondly, different ionization parameters $\xi\equiv
L_X/(n_H\,R^2)$ are expected, with $\xi\approx 10^3$ for ED models and
$\xi\approx 10^5$ for GD models. This would suggest that ED models may
produce a more luminous line (see Ballantyne \& Ramirez-Ruiz 2001),
but smaller ionization parameters can be predicted even for the GD
models if, for example, the emitting material is clumped or the
off-axis burst emission is less intense than the one we receive. The
equivalent widths of the lines may also be very useful. In the ED
models it is impossible to avoid seeing the ionizing continuum
together with the reflected component, while it is natural in GD
models to see only the reflected component. This implies (Ballantyne,
Fabian \& Ross 2002) that a line with EW$>800$~eV can be produced only
by GD models irrespective of the iron abundance.  Finally, broad-band
X-ray spectroscopy could help settle this question. This is because in
the ED case, the line should be due to material newly synthesized by
the exploding star (which is mostly nickel rather than iron; see
Woosley \& Weaver 1995), while in the GD case, the star exploded
several months before the GRB and the nickel has had time to decay
into iron. It is, however, possible for nickel to be bypassed due to
high neutronization (where the iron is directly synthesized by the
exploding star), or by the iron being dragged out from the stellar
core by the expanding fireball.

Measurements of the strength and shape of the reflection spectrum can
yield the geometry, velocity, density and abundances of the scattering
medium. In this paper, we attempt to fit detailed models of such
reflection spectra, computed with the method described by Ross \&
Fabian (1993), to the X-ray afterglow of GRB~991216. This will allow,
for the first time, a self-consistent determination of both the
Fe~K$\alpha$ line and the continuum in a X-ray afterglow of a GRB. A
description of the data and the fitting results are presented in \S
2. In \S 3 we briefly discuss the general constraints that arose from
the fits in the context of both GD and ED models.

\section{Data analysis}
\label{sec:data}

The most significant detection of a X-ray iron line-like feature was
obtained by \textit{Chandra} following the burst GRB~991216.  The
observation lasted 3.4 hours and began $\approx37$~hours after the
initial burst explosion. The afterglow was detected with both the
Advanced CCD Imaging Spectrometer (in spectroscopic mode; ACIS-S) and
with the High Energy Transmission Gratings (HETG). The presence of a
possible \fe\ line in these data was discussed by Piro et al. (2000),
who used the ACIS-S spectrum to estimate the continuum and the higher
resolution gratings spectrum to measure the line properties. In this
section we simultaneously fit the GRB~991216 data from both
detectors with the constant density Compton reflection models of Ross \&
Fabian (1993) (see also Ross, Fabian \& Young 1999).

Following Piro \etal\ (2000), the ACIS-S data between 0.4 and 8~\kev\
(observer's frame) were grouped to have a minimum of 20 counts per
bin, while counts between 1 and 6.2~\kev\ from both the
\begin{figure}
\resizebox{\hsize}{!}{\includegraphics[angle=-90]{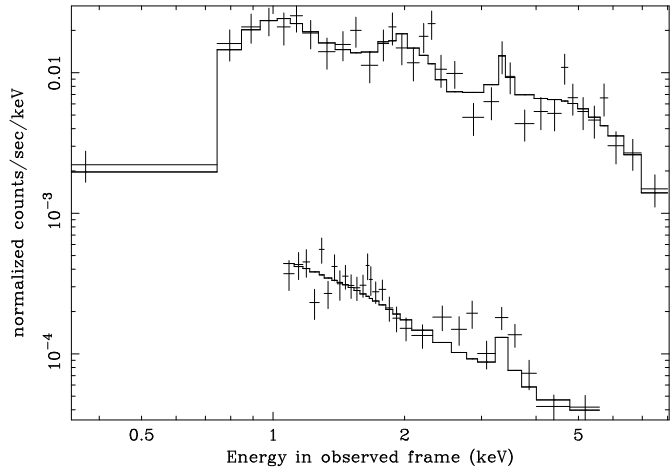}}
\caption{ACIS-S (top) and HETG (bottom) count spectra of the X-ray
afterglow of GRB~991216. The number of counts in the two instruments
are comparable, but the influence of the effective area has been
removed from the HETG data, so it is shown with a smaller
normalized count rate than the ACIS-S spectrum. The best fit Compton
reflection model (\#9; see Table~\ref{tab:summary}) convolved with the
instrumental resolution is shown as solid lines for both spectra.}
\label{fig:spe}
\end{figure}
high-energy and medium-energy gratings were combined to obtain a final
HETG spectrum with 17 counts per bin (see Figure~\ref{fig:spe}). The
slight underbinning of the HETG data does not affect the results. The
normalization of the HETG spectrum was allowed to differ by up to 10\%
from that of the ACIS-S data. A redshift of $z=1.0$ (taken from the
most distant absorption system along the line of sight to the burst;
Vreeswijk \etal\ 2000) was assumed throughout our analysis.  The
spectral fitting was performed using \textsc{xspec}~v.11.0.1aj (Arnaud
1996), and all error bars quoted below are the 90\% confidence limits
for one interesting parameter.

The results of our spectral fitting analysis are summarized in
Table~\ref{tab:summary}.
\begin{table*}
\begin{center}
\begin{minipage}{170mm}
\caption{Summary of fit results. A Galactic absorbing column of
$2.1\times 10^{21}$~cm$^{-2}$ was included in
all fits. In the table heading, $N_{\mathrm{H}}$ denotes the neutral
hydrogen column density (units of 10$^{22}$~cm$^{-2}$) at the redshift
of the source, $E$ is the rest-frame centroid energy (in \kev) of the
Gaussian emission line, $\sigma$ is the width of the line in keV, and `Sig.' denotes the significance of the
added model parameter as given by the F-test. PL=power-law,
A=absorption, R=ionized reflection model \& G=Gaussian.}
\label{tab:summary}
\begin{tabular}{@{}llllllllll}
Fit \# & Model & $\Gamma_1$ & $N_{\mathrm{H}}$ & $E$ & $\sigma$ & $\Gamma_2$ & $\chi^2/$d.o.f. & Sig. & Notes \\ \hline
1 & PL & 1.76$^{+0.09}_{-0.10}$ & -- & -- & -- & -- & 99/60 & -- & -- \\
2 & A*PL & 1.83$^{+0.14}_{-0.13}$ & 0.24$^{+0.40}_{-0.24}$ & -- & -- &
-- &
97/59 & 74\% & -- \\
3$^{\dag}$ & PL & 1.83$\pm 0.11$ & -- & -- & -- & -- & 85/50 & -- & -- \\
4$^{\dag}$ & A*PL & 2.03$^{+0.11}_{-0.19}$ & 0.52$^{+0.55}_{-0.40}$ &
-- & -- & -- &
80/49 & 92\% & -- \\
5$^{\dag}$ & PL+G & 1.87$\pm 0.11$ & -- & 6.80$^{+0.02}_{-0.19}$ &
0.01$^{\mathrm{fixed}}$ & -- & 72/48 & 98\% &  -- \\
6$^{\dag}$ & A*(PL+G) & 2.13$^{+0.23}_{-0.21}$ &
0.63$^{+0.58}_{-0.43}$ & 6.80$^{+0.02}_{-0.19}$ &
0.01$^{\mathrm{fixed}}$ & -- & 65/47 & 97\% & EW $\approx$ 300~eV \\
7$^{\dag}$ & A*(PL+G) & 2.17$^{+0.26}_{-0.21}$ &
0.68$^{+0.63}_{-0.44}$ & 6.82$^{+0.10}_{-0.11}$ &
0.15$^{+0.15}_{-0.11}$ & -- & 62/46 & 84\% & EW
$\approx$ 400~eV \\
8 & same as 6 & -- & -- & -- & -- & -- & 96/57 & -- & -- \\
9 & A*(Bkn. PL+G) & 2.11$^{+0.27}_{-0.19}$ & 0.59$^{+0.63}_{-0.40}$ &
6.80$^{+0.02}_{-0.19}$ & 0.01$^{\mathrm{fixed}}$ &
0.40$^{+0.70}_{-0.84}$ & 72/55 & -- & $E_{\mathrm{brk}}=4.2\pm 0.7$~keV \\ 
10 & A*(PL+R) & 2.61$^{+0.59}_{-0.40}$ & 1.37$^{+0.84}_{-0.51}$
& -- & -- & 0.0$^{+0.44}_{-0.0p}$ & 76/56 & -- & $\log \xi=3.01^{+0.07}_{-0.11}$ \\ \hline
\end{tabular}

\medskip
$_p$ Parameter pegged at lower-limit.
$^{\dag}$ Ignored $E>4.6$~\kev.

\end{minipage}
\end{center}
\end{table*}
A power-law fit to the entire energy range (Fit \#1) does not result
in an adequate fit. Introducing intrinsic absorption (the Galactic
column of $2.1\times 10^{21}$~cm$^{-2}$ is included in all fits) does
not improve this result. The poor fit is a result of a line feature
around 3.4~\kev\ (observed frame) and of a significantly hardening of
the data above $\sim 5$~\kev. This was noticed by Piro \etal\ (2000)
who interpreted the excess above 5~\kev\ as an iron recombination edge
in emission.  Ignoring this change of slope for the time being, we
concentrate on fitting the data below 4.6~\kev\ (the fits marked with
a $\dag$). Fitting a power-law to this limited energy range still does
not provide a good fit to the data (\#3). Adding intrinsic absorption
to the model (\#4) only marginally helps the fit, with the added
component only significant at the 92\% level, according to the
F-test. Replacing the absorber with a Gaussian emission line (Fit \#5)
does improve $\chi^2$ substantially. The line is significant at the
98\% level, and its centroid energy is tightly constrained around the
energy of the He-like \fe\ line\footnote{However, if we ignore the
change in continuum slope and fit a power-law plus Gaussian line model
to the entire energy range, the upper-limit of the line energy is
6.98~keV at $z=1.02$ (roughly the error on the measured
redshift). Therefore, we cannot strictly rule out a contribution from
H-like iron.}.  Including intrinsic absorption in this model (Fit \#6)
improves the fit still further. Allowing the width of the Gaussian to
vary (Fit \#7) does not significantly improve the fit, but does show
that the line is marginally resolved and is likely
broadened. Therefore, the true error on the centroid energy is $\pm
0.1$~\kev, as the narrow line fit above missed a second minimum in
$\chi^2$ space at 6.9~\kev. In summary, we can find an adequate fit to
the data below 4.6~keV with a power-law plus Gaussian emission line
model. The fit also seems to prefer absorption greater than that
provided by the Galactic column, although this is significant only at
the 97\% confidence level.

\begin{figure}
\resizebox{\hsize}{!}{\includegraphics[angle=-90]{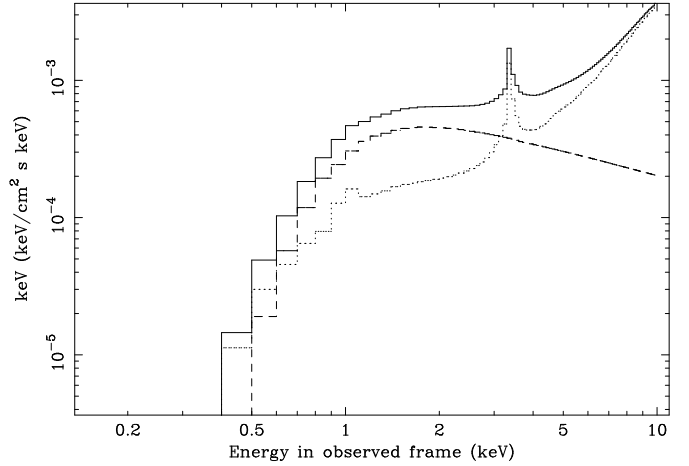}}
\caption{Best fit model (\#10) in physical units.
The solid line show the total spectrum, while the dashed and 
dotted lines show the blast wave emission and reflected component,
respectively.}
\label{fig:phy}
\end{figure}

Reintroducing the harder data, but leaving the model parameters the
same as in Fit \#6, greatly increases the $\chi^2$ of the fit
(\#8). The residuals show that there is a clear change in slope at
higher energies. Fitting the complete continuum with a broken
power-law (\#9) does a very good job in fitting the entire dataset,
and implies that the observed spectrum may be a mixture of two
separate emission features. Replacing the broken power-law model with
a power-law plus Compton reflection spectrum (with incident power-law
included) gives a very similar fit (\#10; Fig.~\ref{fig:phy}).

The photon-indices of the two power-laws are very different, with the
Thomson-thick reflector subject to a very hard ($\Gamma \approx 0$)
ionizing continuum. The ionization parameter of the slab is well
constrained, and, similar to the Gaussian line model, shows that
emission from He-like iron is required with strong
significance. However, the iron abundance and relative strength of the
reflection features cannot be constrained with these data. The above
result was obtained assuming a solar abundance of iron and a reflection
fraction of unity.

\section{Discussion}
\label{sec:discuss}

In this paper we have, for the first time, fitted a physically sound
model to the spectrum of the X-ray afterglow of a GRB that shows
evidence of iron line emission. Our analysis applies to reflection
from an optically thick, homogeneous medium, which seems to be
required in order to explain the large equivalent width that is
observed. The emission feature observed in the X-ray
spectrum of GRB~991216 can be explained by reflection if identified with the
recombination K$\alpha$ line from He-like iron at 6.7~keV. Although
there exists a significant zone of H-like Fe in the reflector, its
K$\alpha$ line is subject to resonant trapping and is ultimately destroyed by
Thomson scattering. If, however, the emitting medium has a velocity
dispersion of $\sim 0.1\,c$ (Piro et al. 2000), resonant trapping
would be suppressed (Lazzati et al. 2001), making the H-like line at
6.97~keV visible. On the other hand, the spectrum can be successfully
modelled without kinematic line broadening since the observed width
can be reproduced by Comptonization for an expansion velocity below
the limit of 0.1$c$ inferred by Piro et al. (2000).

In order to fit the hardening of the spectrum above 5 keV (which was
interpreted by Piro et al. (2000) as a recombination edge) we require
the ionizing continuum to have a much harder power-law than the
observed one. This hard continuum may result from the prompt or
central engine emission and would be primarily responsible for
ionizing the slab. The other, softer spectrum would then originate
from the blast-wave emission, which at later times dominates most of
the observed spectrum.  If this power-law spectrum continues to
$h\nu\ge511$~keV, a significant fraction of the energy in this early
ionizing spectrum will be above the $\gamma\gamma \to e^{\pm}$
formation energy threshold. This will cause new pairs to be formed in
the originally optically thick scattering medium, an effect which
amplifies the density of scattering charges and increases the
temperature of the illuminating material. The effect of pair
production on the line luminosity is twofold. On one hand, the
electron density is increased, making the recombination time shorter
and then the line more luminous. On the other hand, the Thomson
opacity of the slab is increased, reducing the depth out of which line
photons can escape without being scattered by free electrons. These
two effects compensate and as a net result the line luminosity is
slightly decreased by the increase of the electron temperature
(Kallman et al. 2002).  The results presented here are computed only
up to an energy of 100~keV, so we do not take into account pair
processes.

We find that iron enrichment is not necessary to reproduce the line
strength, although if the true ionization parameter is substantially
different than the best-fit value $\xi \approx 10^3$, then a
moderately super-solar abundance will help bring the line to the
required power (Ballantyne \& Ramirez-Ruiz 2001; Ballantyne et
al. 2002). In addition, even if at a non-statistically significant
level, the model seems to underestimate the line strength in the
grating spectrum, and a supersolar metallicity may cure this. Light
elements such as S, Ar and Ca are not currently included in this reflection
model (Ross \& Fabian 1993), so it is not possible to determine if the
small deviation at $h\nu\sim1.8$~keV is consistent with a S
recombination edge.

The value of $\xi \approx 10^3$ is predicted by the extreme clumping
of ED models, but can also be accommodated in GD models if the
reprocessing material is moderately clumped and/or the off-axis burst
ionizing continuum is substantially dimmer than the observed burst
emission. This is not unexpected, since the beaming break in
GRB~991216 occurred quite early (1.5 days; see Halpern et
al. 2000). Unfortunately, the statistical quality of the data does not
allow clear conclusions to be drawn on the line emission mechanism,
and both GD and ED scenarios are consistent with the data.

In summary, our main conclusions are as follows. The iron line-like
feature in the X-ray afterglow of GRB~991216 has properties consistent
with reflection from an optically thick slab with solar metallicity
and ionization parameter $\xi \approx 10^3$. The data constrain the
energy of the line to be $\approx 6.8$~\kev, requiring recombination
emission from He-like iron. There is some evidence for substantial
intrinsic absorption at the source, consistent with the burst
occurring within a gas-rich environment (such as a star-forming
region). No kinematic broadening from an outflow is required by the
data. The ionizing spectrum that impinges on the slab must however be
harder than the afterglow spectrum that dominates the emission at
$h\nu<4$~keV. All these findings can be accounted for by many of the
emission mechanisms and geometries discussed in the
literature. However, we have shown that ionized reflection models can
describe the X-ray afterglow of GRB~991216, and that they may provide
important information on the immediate environments of other bursts.
Higher signal-to-noise data are required in order to have more insight
in the properties and geometry of the line emitting material.

\begin{acknowledgements}
The authors thank the referee for his comments on the manuscript. DRB
acknowledges financial support from the Commonwealth Scholarship and
Fellowship Plan and the Natural Sciences and Engineering Council of
Canada. 
\end{acknowledgements}

\end{document}